\documentclass[fleqn,12pt,twoside]{article}
\usepackage{espcrc1}
\usepackage{graphicx}
\usepackage[figuresright]{rotating}
\usepackage{dcolumn}
\usepackage{bm}
\usepackage{color}
\usepackage{latexsym}
\title{Nuclear structure far from stability}
\author{D. Vretenar
\address{Physics Department, Faculty of Science, University of
Zagreb, Croatia}}
\begin{document}
\maketitle
\begin{abstract}
Modern nuclear structure theory is rapidly evolving 
towards regions of exotic short-lived nuclei far from stability, 
nuclear astrophysics applications, and bridging the gap 
between low-energy QCD and the phenomenology of finite nuclei. The 
principal objective is to build a consistent microscopic theoretical 
framework that will provide a unified description of bulk properties, 
nuclear excitations and reactions. 
Stringent constraints on the microscopic approach to nuclear dynamics, 
effective nuclear interactions, and nuclear energy density functionals, 
are obtained from studies of the structure and stability of exotic nuclei 
with extreme isospin values, as well as extended asymmetric nucleonic matter. 
Recent theoretical advances in the description of structure 
phenomena in exotic nuclei far from stability are reviewed.  
\end{abstract}  

\section{\label{secI}Introduction}
Experimental and theoretical studies of nuclei far from stability 
are at the forefront of modern nuclear science. 
Prompted by a wealth of new experimental data on exotic nuclei with
extreme isospin values, by the rich astrophysical phenomenology, 
as well as theoretical developments in
related fields, important qualitative and 
quantitative advances in theoretical nuclear structure have recently been
reported. Modern nuclear structure theory 
is rapidly evolving from macroscopic and 
microscopic models of stable nuclei towards regions of short-lived nuclei 
close to the particle drip lines. Accurate global 
microscopic calculations have become standard in  
astrophysical applications, and a series of studies based 
on concepts of effective field theory has been initiated in order to 
bridge the gap between low-energy non-perturbative  
QCD and nuclear many-body dynamics.

In light systems quantum Monte Carlo calculations have been extended 
to $A \leq 12$ nuclei and to neutron droplets, and the {\it ab initio}
no-core shell model approach currently 
provides a quantitative description of $p$-shell nuclei.
Large-scale shell model calculations 
have been performed for medium-heavy and heavy nuclei, and applied to 
problems relevant to nuclear astrophysics. In the microscopic description 
of weakly bound neutron-rich nuclei, improved shell model techniques 
allow for a consistent treatment of bound states, resonances and the 
non-resonant continuum background. 

Global shell-model approaches and 
microscopic self-consistent mean-field models have been very successful 
in the description of the evolution of shell structure, the disappearance 
of spherical magic numbers, deformations and shape coexistence in exotic 
nuclei. The evolution of quadrupole collectivity and the coexistence of 
shapes have been analyzed with self-consistent models that include 
correlations beyond the mean-field approximation. Microscopic mass formulas 
based on the self-consistent HFB framework have been developed. The covariant 
mean-field framework has been extended with explicit density-dependent 
effective interactions, which provide an improved description of asymmetric 
nuclear matter, exotic nuclei, hypernuclei, and neutron star matter. 

New theoretical tools have been developed to describe the multipole 
response of neutron-rich nuclei. Several implementations of the (continuum) 
non-relativistic and relativistic quasiparticle random-phase approximation,
as well as the shell model and the quasiparticle phonon model, 
have been employed in studies of the evolution of the low-energy dipole 
and quadrupole response in nuclei with a large neutron excess.

\section{\label{secII}$\bm{Ab~initio}$ and global 
shell-model description of light nuclei}

Light and medium-light nuclei play a particularly important role 
in modern nuclear structure. Experimentally, these nuclei are 
accessible from the proton to the neutron drip line and, therefore, 
provide information on systems with extreme N/Z ratios. Their structure
can be analyzed with a variety of theoretical approaches, including 
exact {\it ab initio} calculations with NN and NNN bare interactions.
Neutron-rich light nuclei exhibit very interesting 
structure phenomena, such as the weak binding of the
outermost neutrons, pronounced effects of the coupling between
bound states and the particle continuum, regions of nuclei with
very diffuse neutron densities, formation of 
halo structures. 

In the past few years a number of microscopic studies have shown
that accurate predictions about the stability, structure and 
reactions of light nuclei can be made starting from the interactions 
among individual nucleons. Energies of all the bound and narrow states
of up to ten nucleons can be reproduced almost exactly (within 2\%)
by employing bare nuclear forces that fit NN scattering data, with 
the addition of realistic NNN forces. The most accurate {\it ab initio}
calculations of ground states and low-lying excitations of light nuclei,
starting from realistic models of the nuclear force, use the 
Quantum Monte Carlo (QMC) method~\cite{PW.01}. 
Most of the QMC calculations have been performed using the 
Argonne $v_{18}$ (AV18) NN potential \cite{AV18}, alone or 
with the inclusion of NNN potentials. The AV18 is representative of the 
modern NN potentials that give accurate fits to scattering data. 
However, with the exception of $^2$H, NN potentials alone cannot 
describe the structure of light nuclei. It has been known for a long
time that Hamiltonians containing only realistic NN potentials 
underbind the light nuclei, overestimate the equilibrium density
of nuclear matter, and cannot reproduce the empirical energy 
spacings between spin-orbit partner levels.  
Already three- and four-nucleon systems provide
ample evidence for the presence of NNN interactions in nuclei.
In addition to increasing the total binding energy, the inclusion 
of NNN interactions improves the level ordering and level spacing 
among low-energy states in comparison with experimental spectra and, 
especially important, enhances the spin-orbit effects. In contrast 
to the NN interaction, however, a detailed form and parameterization 
of the NNN forces are not well established.  
QMC nuclear structure calculations have recently been 
used to construct a set of 
improved pion-exchange NNN potentials, designated the Illinois 
models~\cite{PPWC.01}, by fitting the energies of all the 17 
bound or narrow states of $3 \leq A \leq 8$ nuclei. Used in 
conjuction with the AV18 NN potential, the new Illinois NNN models 
have been very successfully employed in QMC calculations of 
ground and low-lying excited states of $A=9,10$ nuclei~\cite{PVW.02}, 
and of seven- and eight-body neutron drops in 
external potential wells~\cite{PPWC.01}. 
QMC calculations of the 
ground state of fourteen neutrons in a periodic box, approximating 
uniform neutron matter, have recently been reported at densities 
up to one and half times the nuclear matter density~\cite{CMPR.03}. 
In general, however, the computational effort 
increases exponentially with the number 
of nucleons, and with the present computing resources $^{12}$C 
may be the largest symmetric nucleus that can be calculated in the QMC 
framework. A new approach, the auxiliary field diffusion Monte Carlo
method, has been applied to large neutron systems at zero 
temperature ($\approx$ 60 neutrons) \cite{FSS.01}, and should be 
capable of describing nuclei with $A > 12$.

An alternative, complementary approach is the large-basis no-core 
shell model (NCSM). In this {\it ab initio} method the effective 
Hamiltonian is derived microscopically from realistic NN interactions 
as a function of the finite harmonic oscillator basis space.
NCSM calculations have been performed for both $s$-shell and 
$p$-shell nuclei in large, multi-$\hbar \Omega$ model spaces. 
The Argonne and CD-Bonn NN potentials have been used 
in the NCSM calculations of binding energies, excitation spectra, 
electromagnetic properties and Gamow-Teller transitions of $A=10$
nuclei~\cite{CNOV.02}. However, as in the case of QMC calculations, 
the comparison with experimental data indicates the need for 
NNN forces. The NCSM has very recently been extended to include 
effective \cite{NO.02} and realistic \cite{NO.03} NNN interactions 
in calculations of $p$-shell nuclei. The first applications, using 
the Argonne V8' NN potential and the Tucson-Melbourne TM'(99) 
NNN interaction, include $p$-shell nuclei up to $^{13}$C. The inclusion 
of the NNN interaction increases the calculated binding energies and, 
in general, the low-lying spectra are in better agreement with 
experiment. In particular, with the realistic NNN force, the 
correct ground-state spins are obtained for $^{10,11,12}$B and 
$^{12}$N, contrary to calculations with NN potentials only.
An exciting new development is the recent NCSM calculation of 
$^6$Li and $^{10}$B \cite{NC.04}, using the new, effective field 
theory (EFT) based, momentum space nonlocal NN potential at the fourth 
order of the chiral perturbation theory ($N^3LO$) \cite{EM.03}. 
The results are consistent with those obtained with standard 
NN potentials, and identify the need for NNN terms that appear 
already at the third order of the chiral perturbation theory. 

For medium-light nuclei with $A > 12$ 
the method of choice, when applicable, is the global shell 
model approach. The building blocks of the nuclear shell model:
the universal effective interactions, a comprehensive 
treatment of the valence space, and the solution of the secular 
problem in a finite space, have reached a high level of 
sophistication and accuracy \cite{CMNPZ.04}. It is now possible 
to diagonalize matrices in determinantal spaces of dimension 
around $10^9$ using the Lanczos method. New effective interactions 
have been constructed that are connected with both the NN and 
NNN bare forces. The two-body effective interactions are related 
to realistic NN potentials that fit scattering data, 
whereas three-body contributions correct the bad saturation 
and shell-formation properties of realistic two-body forces.
It has been shown that with the inclusion of a simple three-body
monopole Hamiltonian, large-scale shell-model calculations achieve 
a very accurate description of low-energy spectra in the 
$p$, $sd$ and $pf$ shells~\cite{Zuk.03,PSCN.01,HOBM.02}.

The microscopic description of weakly bound and unbound nuclei 
necessitates a consistent treatment of both the many-body 
correlations and the continuum of positive energy states and 
decay channels. The nuclear shell model has recently been 
extended to allow a treatment of an arbitrary number of 
valence nucleons occupying the bound states and the particle 
continuum. The Gamow Shell Model (GSM) \cite{MNPB.02} has been 
formulated using a complex Berggren ensemble representing 
bound single-particle states, single-particle resonances, 
and non-resonant continuum states. The model has been successfully 
tested in calculations involving several valence neutrons 
outside the doubly-magic core: $^{6-10}$He and 
$^{18-22}$O \cite{MNPO.03}, and in the description of $^{5-11}$Li 
including the model space of proton and neutron states \cite{MNPR.04}.
The first results for binding energies, excitation spectra, and 
electromagnetic properties look very promising.  
It has been demonstrated that the contribution of the non-resonant
continuum is crucial, especially for unbound and near-threshold 
states. In some cases (e.g., $^{8,9}$He ) non-resonant continuum 
components dominate the structure of the ground-state wave 
function. In all cases considered, the GSM calculations 
yield neutron resonances above the calculated neutron threshold -- 
a feature that is not imposed {\it a priori} on the model.
In contrast to the standard shell model, the effective interactions 
of GSM cannot be represented as a single matrix calculated for 
all nuclei in a given mass region. The matrix elements that involve
continuum states are strongly system-dependent and they take 
into account the spatial extension of the single-nucleon wave 
functions. For future applications it will be important to develop 
realistic effective interactions to be used in the GSM. It should 
also be emphasized that the dimension of the non-hermitian Hamiltonian
matrix of the GSM grows extremely fast with increasing size of the
Hilbert space, and therefore for a successful application of the GSM 
to heavier nuclei the model basis must be optimized. One promising 
approach is the implementation of a method based on the density 
matrix renormalization group \cite{MNPR.04,DPDS.02}.

\section{\label{secIII}Evolution of shell structure}

The phenomenon of shell evolution in exotic nuclei has been the 
subject of extensive experimental and theoretical studies. Far 
from the $\beta$-stability line the energy spacings between 
single-particle levels change considerably with the number of 
neutrons and/or protons. This can result in reduced spherical shell gaps, 
modifications of shell structure, and in some cases 
spherical magic numbers may disappear. For example, in neutron-rich 
nuclei $N=6,16,34...$ can become magic numbers, while $N=8,20,28...$ 
disappear. The reduction of a spherical shell closure is 
associated with the occurrence of deformed ground states and, in a
number of cases, with the phenomenon of shape coexistence. For 
particular isotopic chains the onset of deformation could extend 
the neutron drip line far beyond the limit expected for 
spherical shapes. 

Both the shell model approach and the self-consistent mean-field models 
have been employed in the description of shell evolution far from 
stability. The basic advantage of the shell model is the ability to
describe simultaneously all spectroscopic properties of low-lying 
states for a large domain of nuclei. Advances in parallel computer 
technology, algorithms and computer codes have extended the range of nuclei 
amenable to a shell model description. Present capabilities include 
all nuclei in the $pf$-shell and the $f_{5/2}$, $p_{3/2}$, $p_{1/2}$, $g_{9/2}$
valence space, as well as heavy semi-magic nuclei, for instance the 
$N=126$ isotones. The region of deformed nuclei around $N=Z=40$ requires
a model space that is prohibitively large for diagonalization shell-model 
approaches, but is feasible in modern stochastic, e.g. Monte Carlo, 
versions of the shell model. 

The origin of the shell evolution and new magic numbers in light 
exotic nuclei has been attributed to the spin-isospin dependent 
central part of the effective NN interaction in nuclei~\cite{OFU.01}.
Although the importance of the $p-n$ $j_> - j_<$ monopole interaction
for the evolution of magicity is still the subject of some 
debate~\cite{ZO.03}, and it seems that very recent experimental data 
do not show evidence for some predicted magic numbers in heavier 
systems (e.g. $N=34$ \cite{Lid.04}), nevertheless 
it has been shown that a shell-model 
Hamiltonian with enhanced spin-flip proton-neutron interaction 
provides an improved description of Gamow-Teller transitions
and magnetic moments in $p$-shell nuclei~\cite{SFO.03}. 
The spectroscopy of proton-deficient nuclei in the $sd$ and 
$pf$-shells provides ample evidence that the traditional magic 
numbers do not extend far from stability. Extensive shell-model and
mean-field studies have predicted the erosion of the spherical 
$N=20$ and $N=28$ shell closures in neutron-rich nuclei.
The results of large-scale shell-model calculations 
are in very good agreement with the recently determined level schemes
of $^{40,42,44}$S \cite{Soh.02}, confirming the predicted onset 
of quadrupole deformation below $^{48}$Ca. The shell model has  
also been applied in the calculation of charge isotope shifts of 
even and odd Ca isotopes \cite{CLM.01}. The model reproduces the 
characteristic features of the isotope shifts, the parabolic dependence
on the mass number and the pronounced odd-even staggering, related to 
the partial breakdown of the $Z=20$ shell closure.

The magicity of the $N=40$ shell and, in particular, the possible 
doubly magic character of $^{68}$Ni has recently attracted considerable
interest. The behavior of the $B(E2, 0^+ \rightarrow 2^+)$ values in the 
Ni isotopic chain illustrates the structural evolution from the doubly 
magic nucleus $^{56}$Ni to $^{68}$Ni \cite{Sor.02}. Although the latter 
nucleus does not display a pronounced discontinuity in the two-neutron
separation energy, the low-lying $0_2^+$ level and the marked 
decrease of the $B(E2, 0^+ \rightarrow 2^+)$ have been interpreted as
evidence for magicity at $N=40$. However, recent microscopic calculations 
of the B(E2) strength distribution in even-even Ni isotopes, using the 
shell model Monte Carlo, the quasiparticle random-phase approximation, 
and a large-scale diagonalization shell model, have shown that in 
$^{68}$Ni the calculated B(E2) value to the first $2^+$ state exhausts only 
a fraction of the low-lying B(E2) strength, and that the 
small experimental B(E2) value to the first $2^+$ state is not a strong 
evidence for the doubly-magic character of $^{68}$Ni \cite{LTN.03}.

The proton-rich deformed nuclei in the $A \approx 80$ mass-region are 
still beyond the capabilities of the shell-model diagonalization approach.
Nevertheless, first shell model Monte Carlo calculations for proton-rich 
Kr, Sr and Zr isotopes in the mass range $A=72-84$ have been 
reported \cite{LDN.03}. By employing the complete $0f1p$-$0g1d2s$ 
configuration space, the calculations reproduce the large empirical 
B(E2) values, and attribute the ground state deformations to the gain in 
the correlation energy obtained by promoting nucleons across the 
$N=40$ subshell closure. In heavy systems 
the evolution of proton shell structure beyond
$^{208}$Pb is of decisive importance for the shell stabilization 
of superheavy elements. Shell model calculations for the $N=126$ 
isotones have been performed for the first time in the 
full proton $Z=82-126$ model space \cite{CRG.03}. In comparison 
with experimental data, excellent results have been obtained for 
binding energies, level schemes and electromagnetic properties. 

Properties of heavy nuclei with a large number of active valence 
nucleons are best described in the framework of self-consistent 
mean-field methods. A broad range of successful applications to 
nuclear structure and low-energy dynamics characterizes mean-field 
models based on the Gogny interaction, the Skyrme energy functional,
and the relativistic meson-exchange effective Lagrangian \cite{BHR.03}.
In recent years important 
advances have been reported in the self-consistent mean-field treatment
of exotic nuclei far from stability. 

A quantitative description of phenomena related to shell evolution 
necessitates the inclusion of many-body effects beyond the mean-field
approximation. The starting point is usually a constrained Hartree-Fock 
plus BCS (HFBCS), or Hartree-Fock-Bogoliubov (HFB) calculation of the 
potential energy surface with the mass quadrupole components as 
constrained quantities. In most applications the calculations have 
been restricted to axially symmetric, parity conserving configurations.
The erosion of spherical shell closures in neutron-rich nuclei produces
deformed intrinsic states and, in some cases, mean-field 
potential energy surfaces with almost degenerate prolate and oblate 
minima. Soft potential energy surfaces and/or
small energy differences between coexisting minima point to the  
importance of including correlation effects. The rotational 
energy correction, i.e. the energy gained by the restoration of 
rotational symmetry, is proportional to the quadrupole deformation of 
the intrinsic state and can reach few MeV for a well deformed configuration.
Fluctuations of the quadrupole deformation also contribute to the 
correlation energy. Both types of correlations can be included simultaneously
by mixing angular momentum projected states corresponding to 
different quadrupole moments. Configuration mixing is usually 
performed by using the generator coordinate method (GCM) with the 
quadrupole deformation as generator coordinate. 
 
In a series of recent papers \cite{RER.02a,RER.02b,RER.03,RER.04}, 
the angular momentum projected GCM with the axial quadrupole moment as 
the generating coordinate, and intrinsic configurations calculated  
in the HFB model with the finite range Gogny interaction, has been applied
in studies of shape-coexistence phenomena that result from the erosion 
of the $N=20$ and $N=28$ spherical shells in neutron-rich nuclei. 
Based on the well known D1S parameterization of the effective Gogny 
interaction, the calculations are completely parameter-free. 
Good agreement with experimental data has been obtained for the $2^+$ 
excitation energies and B(E2) transition probabilities of the 
$N=28$ neutron-rich isotones, and coexistence of shapes has been predicted in
$^{42}$Si, $^{44}$S, and $^{46}$Ar \cite{RER.02a}. The systematic study of 
the ground and low-lying excited states of the 
even-even $^{20-40}$Mg \cite{RER.02b} is particularly interesting, because
this chain of isotopes includes three spherical magic numbers 
$N=8, 20, 28$. It has been shown that the $N=8$ shell closure is 
preserved, whereas deformed ground states are calculated for $N=20$ and 
$N=28$. In particular, the ground state of $^{32}$Mg becomes deformed 
only after the inclusion of the rotational energy correction. The 
$0^+$ collective wave function displays significant mixing of 
oblate and prolate configurations. The deformed ground state of 
$^{32}$Mg occurs as a result of a fine balance between the zero-point 
correction associated with the restoration of rotational symmetry and
the correlations induced by quadrupole fluctuations. In a similar  
analysis of the chain of even-even isotopes $^{20-34}$Ne \cite{RER.03},
it has been found that the ground state of the $N=20$ nucleus $^{30}$Ne 
is deformed, but less than the ground state of its isotone $^{32}$Mg. 
The model has recently been applied in an analysis of the shape 
coexistence and quadrupole collectivity in the neutron-deficient 
Pb isotopes \cite{RER.04}. 
A good qualitative agreement with available data has been
found, especially for the rotational bands built on coexisting 
low-lying oblate and prolate states.
\begin{figure}[htb]
\vspace{-0cm}
\includegraphics[scale=1.0]{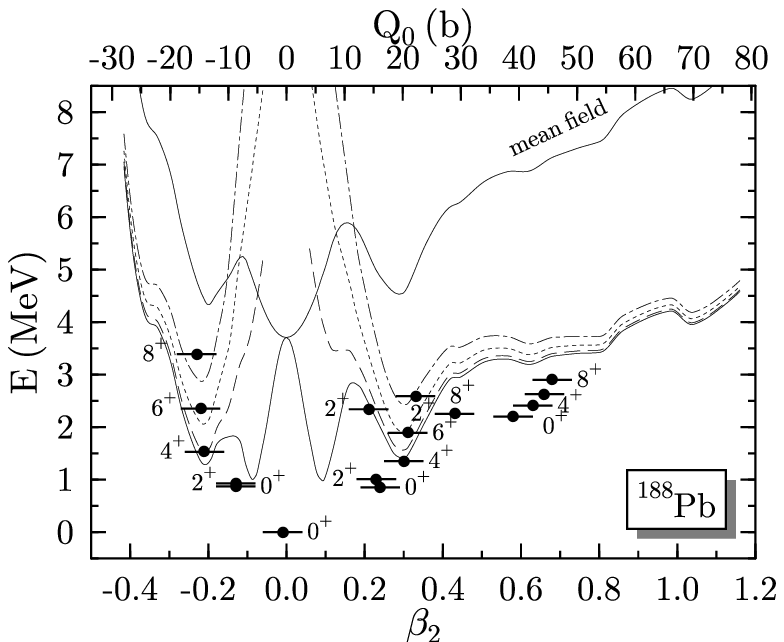}
\includegraphics[scale=1.1]{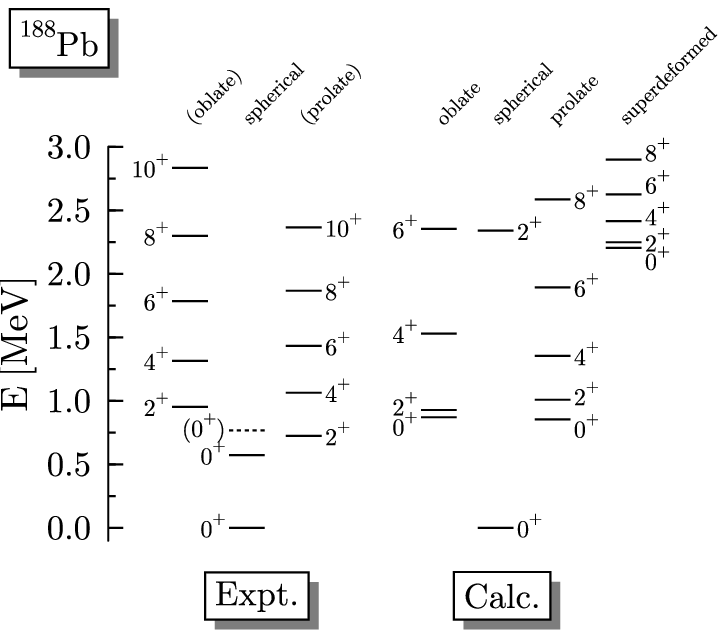}
\vspace{-1cm}
\caption{Particle-number projected (mean-field), particle-number 
and angular-momentum projected potential energy curves up to 
$J=8^+$, and the corresponding lowest GCM states for $^{188}$Pb as 
functions of the quadrupole deformation (left panel) \protect\cite{BBDH.04}.
In the right panel the calculated excitation energies are compared
with the available experimental data \protect\cite{Dra.03}.}
\label{GCM}
\end{figure}
 
Another very sophisticated model \cite{VHB.00} 
which extends the self-consistent mean-field approach
by including correlations, is based on constrained HF+BCS 
calculations with Skyrme effective interactions in the particle-hole 
channel and a density-dependent contact force in the pairing channel.
Particle number and rotational symmetry are restored by projecting 
the self-consistent mean-field wave functions on the correct numbers 
of neutrons and protons, and on the angular momentum. Finally, a mixing 
of the projected wave functions corresponding to different quadrupole 
moments is performed with a discretized version of the generator 
coordinate method. The model has recently been successfully tested in 
the study of shape coexistence in $^{16}$O \cite{BH.03}, and in the
analysis of the coexistence of spherical, deformed, and superdeformed 
states in $^{32}$S, $^{36}$Ar, $^{38}$Ar and $^{40}$Ca \cite{BFH.03}.
For the doubly-magic nucleus $^{16}$O 
this parameter-free approach provides a very good description of 
those low-spin states that correspond to axially and 
reflection-symmetric shapes, and allows the interpretation of 
their structure in terms of self-consistent $np-nh$ states. 
A very important recent application is the study of low-lying 
collective excitation spectra of the neutron-deficient lead isotopes
$^{182-194}$Pb \cite{DBBH.03,BBDH.04}. A configuration mixing 
of angular-momentum and particle-number projected self-consistent 
mean-field states, calculated with the Skyrme SLy6 effective interaction, 
qualitatively reproduces the coexistence of spherical, oblate, prolate 
and superdeformed prolate structures in neutron-deficient Pb nuclei.  
The results are illustrated in Fig.~\ref{GCM}, where the GCM spectra 
of $^{188}$Pb are compared with the recent experimental data \cite{Dra.03}.
In a shell-model language the excited $0^+$ states are generated
by proton excitations across the $Z=82$ spherical shell gap. The mean-field
oblate minimum is associated with $2p-2h$ proton configurations, and
the prolate one with $4p-4h$ proton intruder states. 

In order to describe pairing correlations in weakly bound nuclei close to 
the neutron drip line, new methods have been developed that improve the 
treatment of the continuum coupling in HFB based
models \cite{Gra.01,Gra.02,BBD.02}.

The self-consistent mean-field framework, extended to take into account 
the most important correlations, provides a detailed microscopic 
description of structure phenomena associated with the shell evolution
in exotic nuclei. When compared to the shell model, 
important advantages of this approach include the 
use of global effective nuclear interactions, the treatment 
of arbitrarily heavy systems including superheavy elements,  
and the intuitive picture of intrinsic shapes. 
Further developments will involve  
additional degrees of freedom as generator coordinates, the description
of odd nuclei, the extension to triaxial shapes, the inclusion of 
negative parity structures, and the use of effective interactions 
that have been readjusted to take into account the explicit 
treatment of correlations.

\section{\label{secIV}Towards a universal energy density functional}
  
The self-consistent mean-field approach to nuclear structure
represents an approximate implementation of Kohn-Sham density 
functional theory (DFT). The DFT
enables a description of the nuclear many-body problem in terms of
a universal energy density functional, and mean-field models
approximate the exact energy functional, which includes all
higher-order correlations, with powers and gradients of
ground-state nucleon densities. Although it models the 
effective interaction between nucleons, a general density functional 
is not necessarily related to any given NN potential.
By employing global effective interactions,
adjusted to reproduce empirical properties of symmetric and asymmetric 
nuclear matter, and bulk properties
of simple, spherical and stable nuclei, the current generation of 
self-consistent mean-field methods has achieved a high level of 
accuracy in the description of ground states and
properties of excited states in arbitrarily heavy nuclei, exotic nuclei
far from $\beta$-stability, and in nuclear systems at the nucleon
drip-lines.

Concerning the predictive power of these methods, however, the 
situation is far from being satisfactory and there are many issues 
to be addressed. For instance, until very 
recently global effective interactions were adjusted to ground-state
properties of no more than ten or so spherical nuclei. Nuclear masses 
calculated with these interactions have a typical rms deviation of 
$\approx 2$ MeV when compared with experimental mass tables. 
Even though calculated one- and two-nucleon separation 
energies for nuclei not far from stability are usually fairly accurate, 
various non-relativistic and relativistic mean-field
models differ significantly in the prediction of separation 
energies close to the drip lines and, in general, of isovector
properties far from stability. It must be also emphasized that 
global effective interactions have not been optimized to go 
beyond the mean-field plus pairing approximation and therefore,  
when used in calculations that explicitly include correlations, 
they might lead to results that are not completely reliable.

One of the major goals of modern nuclear structure is, therefore, 
to build a universal energy density functional theory \cite{LNP.641}. 
Universal in the sense that the same functional is used for all
nuclei, with the same set of parameters. 
This framework should then provide a reliable microscopic description of 
infinite nuclear and neutron matter, ground-state properties of 
all bound nuclei, low-energy vibrations, rotational spectra, 
small-amplitude vibrations, and large-amplitude adiabatic properties.   
The next generation of energy density functionals should move away 
from model dependence by including all terms allowed by symmetries, 
and all available data, rather than a small subset of spherical nuclei,
should be used in adjusting phenomenological parameters. 
Properties of nuclei far from stability, in particular, should 
provide stringent constraints on the isovector channels
of effective nuclear interactions. New interactions in turn will 
enable an improved description of structure phenomena in exotic systems 
and more reliable extrapolations toward the drip lines. 

The first step in this direction is the construction of microscopic 
mass tables. Recent trends in the determination of nuclear masses have 
been reviewed in Ref.~\cite{LPT.03}. From the theoretical point of view 
an important advance is the development of self-consistent 
Skyrme Hartree-Fock (HF) and Skyrme Hartree-Fock-Bogoliubov (HFB) mass 
formulas. In a series of recent papers~\cite{SGHPT.02,SGP.03,GSHPT.02,GSBP.03}
a set of complete microscopic mass tables of more than 9000
nuclei lying between the particle drip lines over the range $Z,N \geq 8$ and 
$Z \leq 120$, have been constructed within the HFB framework. By adjusting 
the parameters of the Skyrme interaction, the strength and the cut-off 
parameters of the (possibly density-dependent) $\delta$-function pairing 
force, and the parameters of two phenomenological Wigner terms, with 
a total of $\approx 20$ parameters in all, the measured masses of 2135 
nuclei with $Z,N \geq 8$ have been fitted with an rms error of 
less than 700 keV. In addition, although these effective interactions 
have been adjusted only to masses, they also produce excellent results 
for the charge radii, with an rms deviation of $\approx 0.025$ fm 
for the absolute charge radii and charge isotope shifts of more than 
500 nuclei \cite{BPG.01}.
However, despite the impressive quality of the 
Skyrme-HFB mass formulas, they are far from being definite, and a number 
of very recent studies have considered possible modifications to the 
interactions, a better treatment of symmetry-breaking effects and 
many-body correlations, and improved methods of calculations. 
In Ref.~\cite{SDNPD.03} an improved version of 
the deformed configuration-space HFB method has been reported, based 
on the expansion of the HFB wave functions in a complete set of 
transformed harmonic-oscillator basis states, obtained by a local-scaling
point transformation. This method enables a careful treatment of the 
asymptotic part of the nucleonic density, and is therefore 
particularly suitable for self-consistent HFB calculations of 
deformed weakly-bound nuclei close to the nucleon drip lines. 
In the first application, the Skyrme force SLy4 and volume pairing 
have been employed in the calculation of the entire deformed even-even 
mass table for $Z\leq 108$ and $N\leq 188$, with exact particle 
number projection following the application of the Lipkin-Nogami 
prescription. 

When considering proton-rich nuclei with $Z \approx N$, 
additional proton-neutron ($pn$) correlations, e.g. $pn$ pairing,
have to be taken into account.  
Even though the effect of these correlations on the binding energies
can be approximated by phenomenological Wigner 
terms \cite{LPT.03,SGHPT.02,SGP.03,GSHPT.02,GSBP.03}, it is important
to extend the self-consistent mean-field methods to incorporate 
proton-neutron mixing. In Ref.~\cite{PRDN.04} the coordinate-space 
HFB framework has been generalized to include arbitrary mixing between 
protons and neutrons both in the particle-hole and particle-particle 
channels. The resulting HFB density matrices have a rich spin-isospin 
structure and provide a microscopic description of pairing correlations
in all isospin channels. 

In global microscopic mass tables one usually treats various mean-field
and pairing effects very carefully, whereas additional correlations,
related to the restoration of broken symmetries and to fluctuations, 
are either neglected or taken into account in a very schematic way. 
Correlations are, however, very important if the goal is a level of
accuracy better than 500 keV. Unfortunately, the standard microscopic 
methods for treating correlations are computationally far too 
expensive when applied in calculations involving thousands of nuclei. 
It will be very useful to develop approximate methods of 
calculating correlations, that would reduce the computational cost
and therefore enable a systematic calculation of correlation energies
for the nuclear mass table~\cite{BBH.04}. 

An important class of self-consistent mean-field models belongs 
to the framework of relativistic mean-field theory (RMF).
RMF-based models have been very successfully employed in 
analyses of a variety of nuclear structure phenomena, 
not only in nuclei along the valley of $\beta$-stability, 
but also in exotic nuclei with extreme isospin 
values and close to the particle drip lines. Applications 
have reached a level of sophistication and 
accuracy comparable to the non-relativistic 
Hartree-Fock (Bogoliubov) approach based on Skyrme or 
Gogny effective interactions. Nevertheless, standard RMF 
Lagrangians with nonlinear meson self-interactions are plagued 
with a number of problems, especially when describing isovector 
properties. Recently it has been shown that much better results 
are obtained in an effective hadron field theory with medium dependent
meson-nucleon vertices. Such an approach retains the basic
structure of the relativistic mean-field framework, 
but can be more directly related to the
underlying microscopic description of nuclear interactions.
One way to determine the functional form of the meson-nucleon 
vertices is from in-medium Dirac-Brueckner interactions, obtained
from realistic free-space NN interactions. This microscopic approach 
represents an {\it ab initio} description of nuclear matter and
finite nuclei. The corresponding density-dependent relativistic 
hadron field model has recently been applied to
asymmetric nuclear matter and exotic nuclei~\cite{HKL.01},
hypernuclei~\cite{KHL.00,KL.02}, and neutron star 
matter~\cite{HKL.01a}. On the other hand, a phenomenological
approach was adopted in Ref.~\cite{TW.99}, where the density dependence  
for the $\sigma$, $\omega$ and $\rho$ meson-nucleon
couplings was adjusted to properties of nuclear matter and 
a set of spherical nuclei.
This phenomenological effective interaction was further improved in
Ref. \cite{NVFR.02}, where the relativistic
Hartree-Bogoliubov (RHB) model was extended to include medium-dependent
vertex functions. 
It has been shown that, in comparison with standard
non-linear meson self-interactions, relativistic models with an
explicit density dependence of the meson-nucleon couplings provide
an improved description of asymmetric nuclear matter, neutron
matter and nuclei far from stability. The relativistic
random-phase approximation (RRPA), based on effective Lagrangians
characterized by density-dependent meson-nucleon vertex functions,
has been derived in Ref. \cite{NVR.02}.  
A comparison of the RRPA results on multipole
giant resonances with experimental data provide 
additional constrains on the parameters
that characterize the isoscalar and isovector channels of the
density-dependent effective interactions.  
In a microscopic analysis of the nuclear matter compressibility
and symmetry energy \cite{VNR.03}, it has been shown that
the experimental data on the giant monopole resonances
restrict the nuclear matter compression
modulus of structure models based on the relativistic
mean-field approximation to $K_{\rm nm}\approx 250 - 270$ MeV,
while the isovector giant dipole resonances and the
available data on differences between neutron and proton radii
limit the range of the nuclear matter symmetry energy at
saturation (volume asymmetry) of these effective interactions 
to 32 MeV $\leq a_4 \leq$ 36 MeV. The RMF framework with minimal
meson-nucleon couplings has recently been generalized by introducing 
couplings of the meson fields to derivatives of the nucleon field 
in the Lagrangian density \cite{TCW.03}. This approach leads
to nucleon self-energies that depend on both density and momentum, 
and enables an effective description of a state-dependent 
in-medium interaction in the mean-field approximation.

\section{\label{secVI}Exotic modes of excitations}
  
The multipole response of unstable nuclei far from the line of $\beta$%
-stability presents a very active field of
research. On the neutron rich side
the modification of the effective nuclear potential leads to
the formation of nuclei with very diffuse neutron densities, and to the
occurrence of the neutron skin and halo structures. These phenomena will
also affect the multipole response of unstable nuclei, in particular the
electric dipole and quadrupole excitations, and new modes of excitations
could arise in nuclei near the drip line. New models have recently 
been developed that describe the low-lying collective excitations in 
weakly bound nuclei by explicitly taking into account the coupling to 
the continuum \cite{HS.01,Mat.01,Khan.02}.

The dipole response of very neutron-rich isotopes is characterized by the
fragmentation of the strength distribution and its spreading into the
low-energy region, and by the mixing of isoscalar and isovector modes. It
appears that in most relatively light nuclei the onset of dipole strength in
the low-energy region is due to non-resonant independent single-particle
excitations of the loosely bound neutrons.
The structure of the low-lying dipole strength, however,
changes with mass and in heavier nuclei low-lying dipole states appear
that are characterized by a more distributed structure of the (Q)RPA
amplitude. Among peaks characterized by single particle transitions,
collective dipole states are identified below 10 MeV, and their
amplitudes represent coherent superpositions of many neutron particle-hole
configurations. The corresponding transition densities reveal the 
dynamics of a pygmy dipole resonance (PDR): the neutron skin oscillates 
against an $N \approx Z$ core.

The dipole response in neutron-rich oxygen isotopes has 
attracted considerable interest \cite{Lei.01,Try.01}.
Low-lying E1 strength has been observed exhausting 
about 10\% of the classical dipole sum rule below 15 MeV excitation
energy. Extensive theoretical studies, including large-scale
shell-model calculations \cite{SS.99}, 
relativistic (Q)RPA \cite{Vre.01,PRNV.03},
and the QRPA plus phonon coupling model \cite{CB.01}, reproduce
the experimentally observed redistribution of the E1 strength 
in neutron-rich oxygen isotopes, but they also show that the 
low-lying dipole states represent a new type of non-resonant independent 
single-particle excitations, not caused by a coherent superposition
of particle-hole ({\it ph}) configurations like in collective states. 
The structure of the QRPA amplitudes of all low-lying dipole states
is dominated by one, or at most two, single-neutron $ph$ excitations.

Very recently experimental data have been
reported on the concentration of electric 
dipole strength below the neutron separation energy in $N=82$
semi-magic nuclei. The distribution of the electric dipole strength in $%
^{138}$Ba, $^{140}$Ce, and $^{144}$Sm displays a resonant structure between
5.5 MeV and 8 MeV, exhausting $\approx$ 1\% of the isovector 
E1 energy weighted sum rule~\cite{Zil.02}. 
In $^{138}$Ba negative parity quantum numbers have been assigned
to 18 dipole excitations between 5.5 MeV and 6.5 MeV~\cite{Pie.02}.
The experimental information available presently, however, does 
not allow one to determine the dominant structure of the observed states.
The Sn isotopes present another very interesting example of the evolution of
the low-lying dipole strength with neutron number. 
The relativistic QRPA has been employed in the analysis of the 
distribution of dipole strength in Sn isotopes 
and N=82 isotones \cite{PRNV.03}. It has been shown that 
in neutron-rich nuclei a relatively strong
peak appears in the dipole response below 10 MeV, with a QRPA amplitude
characterized by a coherent superposition of many neutron quasiparticle
configurations. This collective mode represents 
the oscillation of the skin of excess neutrons against the
core composed of an equal number of protons and neutrons: the 
pygmy dipole resonance. Beyond the QRPA level,
the PDR in neutron-rich Sn nuclei has been investigated by employing the 
Quasiparticle Phonon Model (QPM), which provides a consistent 
unified description of low-energy single- and multi-phonon states. 
QPM calculations, including up to three-phonon configurations, 
have shown that anharmonic admixtures induce a considerable 
fragmentation of the low-lying 
E1 strength \cite{TLS.04}. Nevertheless, despite significant multi-phonon 
contributions, the PDR states retain their basically one-phonon 
character. An important result of the QPM analysis is
that the fragmentation pattern is reduced when increasing the 
neutron excess toward $^{132}$Sn.

Besides being intrinsically interesting as an exotic mode of excitation, 
the occurrence of the PDR has important implications for the 
r-process nucleosynthesis. Namely, although the E1 strength of 
the PDR is small compared to the total dipole strength, if located 
well below the neutron separation energy the PDR can significantly
increase the radiative neutron capture cross section on neutron-rich
nuclei, as shown in recent large-scale QRPA calculations 
of the E1 strength for the whole nuclear chart \cite{GK.02,GKS.04}.
  
Pygmy dipole resonances are not necessarily limited to exotic nuclei
far from stability. This phenomenon can be expected to occur in all 
medium-heavy and heavy systems characterized by a significant  
excess of neutrons. For example, a recent relativistic RPA 
calculation \cite{VPRL.01} had predicted the PDR in $^{208}$Pb
at the neutron emission threshold. The resonance has been subsequently 
observed in a high-resolution ($\gamma, \gamma^\prime$) study 
of the electric dipole response \cite{Rye.02}, and the data have been very
accurately described in a QPM study with realistic model spaces, 
including the coupling of up to three-phonon configurations.
The QPM analysis of the velocity distributions of the E1 excitations 
in $^{208}$Pb has shown that the $1^-$ states close to the particle
threshold contain pronounced components that correspond to vortex 
collective motion. This result is in agreement with the prediction 
of the occurrence of the toroidal dipole resonance in this energy 
region \cite{VPRN.02}. The toroidal dipole mode corresponds to 
a transverse zero-sound wave and its experimental observation invalidates
the hydrodynamical picture of the nuclear medium, since there is no 
restoring force for such modes in an ideal fluid. 
Another exotic E1 mode in the energy region below giant resonances 
has recently been the subject of intensive experimental and 
theoretical investigation. This is the low-energy 
component of the isoscalar dipole resonance (ISGDR) (for a recent 
review of the current experimental work on the ISGDR, 
see Ref.~\cite{Garg.04}). Although in a classical picture the ISGDR 
corresponds to a second order compression mode and therefore provides
information on the nuclear matter compression modulus, several 
theoretical studies \cite{Col.00,VWR.00,Pie.01,SS.02} have predicted the 
existence of a low-energy component that is not sensitive to the 
nuclear compressibility.

Of course, not only dipole states characterize the exotic 
low-energy multipole response of neutron-rich nuclei. 
Many other interesting phenomena have been observed 
and analyzed. For example, in certain 
neutron-rich Te isotopes a decrease in the energy of the first excited 
$2^+$ state is accompanied by a decrease in the corresponding 
$B(E2, 0^+ \rightarrow 2^+)$. This behavior of the E2 strength 
contradicts the simple systematics observed in most isotopic chains, i.e. 
an increase of the E2 strength as the first excited $2^+$ state 
decreases in energy and becomes more collective. In the QRPA analysis
of Ref.~\cite{TENS.02} this anomalous behavior of $2^+$ excitations 
around $^{132}$Sn has been attributed to a reduction in the neutron pairing 
above the $N=82$ spherical shell gap. 
\begin{figure}[t]
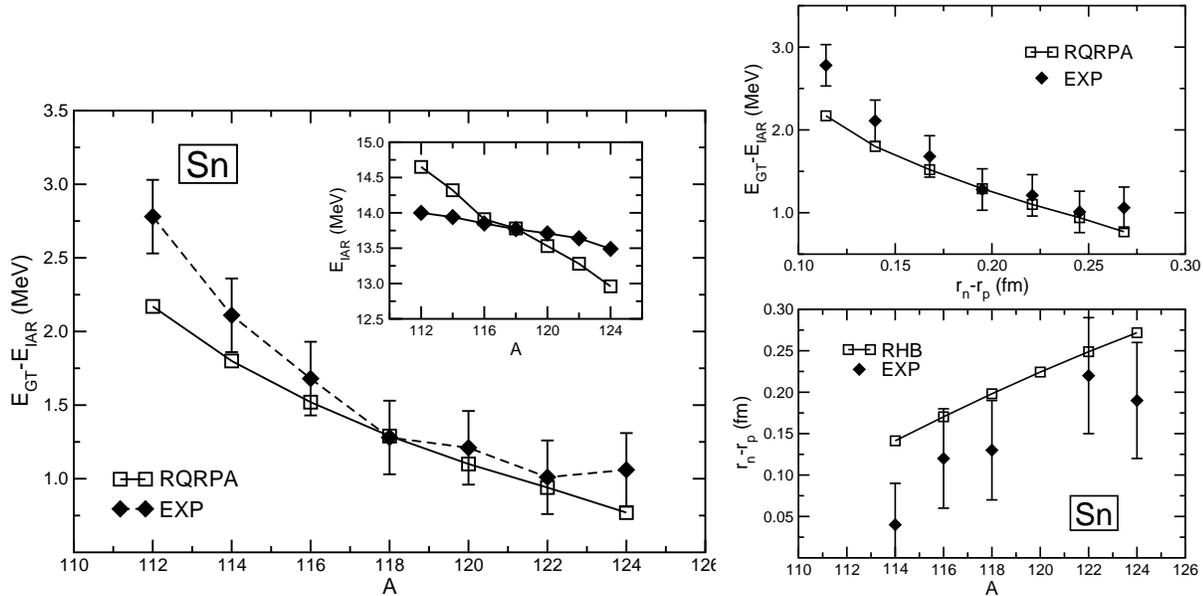

\vspace{-0cm}
\begin{center}
\includegraphics[scale=0.4]{fig1.eps}
\includegraphics[scale=0.35]{fig2.eps}
\end{center}
\vspace{-1cm}
\caption{The proton-neutron RQRPA and experimental
differences between the excitation energies
of the GTR and IAS for the 
sequence of even-even $^{112 - 124}$Sn target nuclei, as a function 
of the mass number (left panel) and of the calculated 
differences between the rms radii of the neutron and proton 
density distributions of even-even Sn isotopes (upper right panel).
In the lower panel on the right the calculated differences $r_n - r_p$
are compared with experimental data \protect\cite{Kra.99}.}
\label{GTR}
\end{figure}

The systematics of low-energy pygmy modes could in principle be used 
to determine the thickness of the neutron skin in neutron-rich 
nuclei \cite{PRNV.03,TLS.04}. Two experimental methods based on
giant resonances have been used to
study the neutron skin: the excitation of the giant
dipole resonances (GDR) \cite{Kra.91} and the spin-dipole 
resonance \cite{Kra.99}. Recently a new method 
has been suggested for determining the difference 
between the radii of the neutron and proton density distributions along 
an isotopic chain, based on measurement of the excitation energies 
of the Gamow-Teller resonance (GTR) relative to the 
isobaric analog state (IAS) \cite{VPNR.03,PNVR.04}. 
The concept is illustrated 
in Fig.~\ref{GTR}, where the results of a relativistic
proton-neutron quasiparticle random-phase
approximation calculation for the excitation energies 
of the main component of the GT 
resonances and the respective isobaric analog states for the 
sequence of even-even $^{112 - 124}$Sn, are shown 
in comparison with data, and the 
calculated differences $r_n - r_p$
are compared with experimental values deduced from the 
excitation of the SDR \cite{Kra.99}. For a given effective nuclear 
interaction which reproduces the ground-state 
and the excitations of the IAS and the GTR, the correlation
shown in Fig.~\ref{GTR} can be used to determine the 
value of the neutron skin. 

The ability to model the Gamow-Teller response is, of course, essential 
for reliable predictions of $\beta$-decay rates in neutron-rich 
nuclei along the $r$-process path. The calculation of GT strength,
however, can also be used to constrain the spin-isospin channel
of energy density functionals. For the Skyrme energy functionals, 
an initial analysis of the couplings of the spin-isospin 
terms bilinear in time-odd densities and currents, has been reported
in Ref.~\cite{BDEN.02}. In the relativistic framework the time-odd 
channels are not independent from the time-even ones because 
they arise from the small components of the Dirac spinor and, 
therefore, a very natural microscopic model for the description of 
spin-isospin excitations is the proton-neutron relativistic QRPA
(PN-RQRPA). In a recent work \cite{PNVR.04}, the PN-RQRPA 
has been formulated in the canonical single-nucleon basis of the 
relativistic Hartree-Bogoliubov model, and employed in the 
analysis of charge-exchange modes in open-shell neutron-rich 
nuclei, with particular emphasis on the role of the $T=1$ and
$T=0$ pairing channels. 

\section{\label{secVII}Microscopic predictions for astrophysical applications}
  

The latest theoretical and 
computational advances in nuclear structure modeling have also 
had a strong impact on nuclear astrophysics. More and more often 
calculations of stellar nucleosynthesis, nuclear aspects of 
supernova collapse and explosion, and neutrino-induced reactions, 
are based on microscopic global predictions for the nuclear 
ingredients, rather than on coarse and oversimplified 
phenomenological approaches. The nuclear input for astrophysics 
calculations necessitates the knowledge of the properties of 
thousands of nuclei far from stability, including the 
characteristics of strong, electromagnetic and weak interaction 
processes. Most of these nuclei, especially on the neutron-rich 
side, are not accessible in experiments and, therefore, many 
nuclear astrophysics calculations crucially depend on accurate 
theoretical predictions for the nuclear masses, bulk properties, 
nuclear excitations, ($n,\gamma$) and ($\gamma,n$) rates, 
$\alpha$- and $\beta$-decay half-lives, fission probabilities, 
electron and neutrino capture rates, etc. 

The path of the $r$-process nucleosynthesis runs through regions 
of very neutron-rich nuclei. $\beta$-decays are particularly 
important because they generate 
elements with higher Z-values, and set the time scale of
the $r$-process. Except for a few key nuclei, however, 
$\beta$-decays of $r$-process nuclei have to be determined 
from nuclear models. Both the shell model and the QRPA 
have been employed in large-scale calculations of weak 
rates for the $r$-process. Shell model applications in
nuclear astrophysics have recently been reviewed in 
Refs.~\cite{LM.02,LM.03}. In the calculation of 
$\beta$-halflives, in particular,
the principal advantage of the shell model is the ability 
to take into account the detailed structure of the $\beta$-strength
functions. In addition to large-scale shell model predictions 
for the half-lives of waiting-point nuclei at $N=50, 80, 126$ \cite{LM.03},
the no-core shell model \cite{CNOV.02} and the shell model 
embedded in the continuum \cite{MONP.02} have been 
applied to $\beta$-decay of light nuclei.
The self-consistent HFB plus continuum QRPA framework has been 
employed in a systematic calculation of 
the allowed and first-forbidden $\beta$-decay rates for the $r$-process 
nuclei near $N=50, 80, 126$ \cite{Bor.03}. It has been shown 
that the effect of the high-energy first-forbidden transitions 
is crucial for $Z\geq 50$, $N\approx 82$, and especially 
in the $N=126$ region.

Improved large-scale shell model results for weak-interaction rates, 
both for beta decay and electron capture, have been used in a study 
of presupernova evolution of massive stars \cite{HLMW.01}. Compared 
to standard presupernova models, larger values of the electron-to-baryon
ratio at the onset of collapse and smaller iron masses have been predicted, 
and these can have significant consequences for nucleosynthesis and
the supernova explosion mechanism. A new model for calculating 
electron capture rates on neutron-rich nuclei has been developed \cite{LKD.01},
based on a combination of shell model Monte Carlo calculation of 
temperature-dependent occupation numbers of single-particle orbitals, 
and RPA calculation of electron capture rates including both allowed 
and forbidden transitions. The model has been used to calculate 
rates of electron capture on nuclei with mass numbers $A=65-112$ for
the temperatures and densities characteristic for core collapse \cite{Lan.03}.
It has been shown that, in contrast to previous assumptions, these 
rates are so large that electron capture on neutron-rich nuclei dominates over 
capture on free protons. With realistic treatment of electron capture 
on heavy nuclei, significant changes in the hydrodynamics of core 
collapse and the properties of the core at bounce have been 
predicted \cite{Hix.03}. 

Neutrino-induced reactions on heavy neutron-rich nuclei in the 
post collapse supernova environment could play an important role 
in the $r$-process. In a very recent calculation of 
neutrino-induced fission cross sections in competition with 
neutron emission~\cite{KLF.04}, it has been shown that 
neutrino charged-current capture-induced fission could 
have a pronounced effect on the nuclear reaction flow paths 
and nuclear abundances in the $r$-process. 
\section{\label{secVIII}
Nuclear structure as an Effective Field Theory of low-energy QCD}
  
The most fundamental problem in
theoretical nuclear physics is: how to establish a relationship
between low-energy, non-perturbative QCD and the rich nuclear phenomenology
which includes both nuclear matter and finite nuclei?
The success of the non-relativistic and 
covariant self-consistent mean-field approach to nuclear
many-body dynamics, and the recent application of chiral effective
field theory to nucleon-nucleon scattering and the few-body problem, 
point to a unified microscopic framework
based on density functional theory (DFT) and effective 
field theories (EFT).

At the energy and momentum scales characteristic of nuclei, 
QCD is realized as a theory of pions coupled to nucleons. 
The basic concept of a low-energy EFT is the separation of scales:
the long-range physics is treated explicitly (e.g. pion exchange) 
and short-distance interactions, that cannot be resolved at low-energy,
are replaced by contact terms. In principle, all interaction terms allowed by
symmetries must be included and the EFT framework, i.e. power
counting for diagrams and gradient expansions, can be used to
remove model dependence. EFT building of the universal
microscopic energy density functional allows error estimates to be
made, and it also provides a power counting scheme which separates
long- and short-distance dynamics. Recent investigations
\cite{HF.00,KFW.01,KFW.02,PBF.03} have shown that models based on EFT and
DFT provide a systematic framework for nuclear structure theory.
A novel microscopic covariant description of nuclear many-body dynamics 
constrained by chiral symmetry and in-medium QCD sum rules has 
been developed \cite{FKVW1,FKVW2}. It has been shown that nuclear binding 
and saturation are essentially generated by chiral (two-pion exchange) 
fluctuations in combination with Pauli effects, whereas strong scalar and 
vector fields of about equal magnitude and opposite sign, induced by changes 
of the QCD vacuum in the presence of baryonic matter, generate the large 
effective spin-orbit potential in finite nuclei. Promising results have been 
obtained for the nuclear matter equation of state and for the bulk and 
single-nucleon properties of finite nuclei. A quantitative 
{\it ab initio} description of the nuclear many-body problem, 
starting from the fundamental theory of QCD, should become feasible 
in the near future.

\end{document}